\documentclass[mathleft,usenatbib]{an}
\usepackage{graphicx}
\usepackage{times}
\usepackage{epsfig}
\usepackage{rotating}
\usepackage{sidecap}
\usepackage{longtable}
\usepackage{lscape}

\begin{document}

% The following seven commands are intended for editorial usage and should be ignored by
% the author(s).
\Pagespan{1}{}% Document's page range. 
% If second parameter is left empty, the last page is computed automatically.
\Yearpublication{2016}%
\Yearsubmission{2015}%
\Month{11}%   
\Volume{999}%  
\Issue{88}% 
% \DOI{This.is/not.aDOI}% 

\title{Quasi periodic oscillations in black hole binaries}

\author{Sara E. Motta\inst{1}\fnmsep\thanks{Corresponding author:
  \email{sara.motta@physics.ox.ac.uk}\newline}
%Example 
%for footnote, note the usage of the \texttt{fnmsep}
%command as separator between institute number and footnote mark} 
}
\titlerunning{QPOs in Black hole X-ray binaries}
\authorrunning{Sara E. Motta}
\institute{
University of Oxford, Department of Physics, Astrophysics,
Denys Wilkinson Building, Keble Road, OX1 3RH, Oxford, United Kingdom
}

\received{...}
\accepted{...}
\publonline{...}


\abstract{%
Fast time variability is the most prominent characteristic of accreting systems and the presence of quasi periodic oscillations (QPOs) is a constant in all accreting systems, from cataclysmic variables to AGNs, passing through black hole and neutron star X-ray binaries and through the enigmatic ultra-luminous X-ray sources. 
In this paper I will briefly review the current knowledge of QPOs in black hole X-ray binaries, mainly focussing on their observed properties, but also mentioning the most important models that have been proposed to explain the origin of QPOs over the last decades.}

\maketitle

\section{Introduction}

X-ray binaries (XRBs) are double systems formed by a stellar remnant that has collapsed to a compact object (typically a black-hole, BH, or a neutron star, NS) and has remained gravitationally bound to its companion. The compact objects attracts the matter on the companion star forcing it to leave its surface. The accretion flow that carries matter from the companion star to the compact object originates a so-called \textquoteleft accretion disk\textquoteright \ around the compact object, where the angular momentum of matter is transported outwards and the gravitational potential energy is converted in kinetic energy and radiation. Accretion disks emit all along the electromagnetic spectrum, even though they are particularly bright in the X-rays, where the radiation coming from the innermost regions of the flow can be observed. 
Here I will focus on black hole XRBs, even though I will mention some similarities that can be found between BH XRBs and NS XRBs when considering their variability and in particular QPOs.

Almost all BH XRBs are transient systems, i.e. they alternate long quiescence phases to short \textquoteleft outbursts\textquoteright, active periods typically lasting from weeks to months. % during which their luminosity rises of several order of magnitudes, showing prominent variability on several timescales. 
Besides being very bright in the X-ray sky, XRBs have a very prominent characteristic: differently from most astrophysical objects, they show significant variability on humanly accessible time scales.  The longest ones (weeks, months, years) can be appreciated inspecting the long term lightcurves (see e.g.  \cite{Dunn2010}). 
These long-term systematic variations, correspond to significant changes in the energy spectra and can be described in terms of a pattern traced in an X-ray hardness-intensity diagram (HID) (see e.g.  \cite{Homan2001} and \cite{Belloni2011}). In most BH candidates, four different canonical spectral/timing states are found to correspond to different branches/areas of a q-shaped HID pattern. Sometimes, a so-called \textit{anomalous state} can also be seen. The analysis of the fast timing variations observed in the sources' power density spectra (PDS) plays a fundamental role in the state classification (see \cite{Homan2005}, \cite{Belloni2010}). %In this scheme LFQPOs, appear to be confined within a 
The states that have been identified in HIDs from many sources (See Belloni and Motta 2015 and \cite{Belloni2011}) are: 
\begin{itemize}
\item the Low Hard State (LHS), where the emission is dominated by a strong Comptonized emission,  occasionally accompanied by a cool and faint thermal component associated to a truncated disk. The intrinsic variability of the sources in this state can reach 30-40\%.

\item the High Soft State (HSS), where the emission is dominated by a strong thermal component, associated to an accretion disk whose radius can extend down to the Innermost stable circular orbit. During this state the variability is very low, often consistent with zero. 

\item the Hard Intermediate State (HIMS) and the Soft Intermediate State (SIMS). During these two states the spectrum presents both a significant hard component and the contribution of a thermal disk. The variability can vary a lot and normally ranges between 5 and 20\%. 

\item the \textit{anomalous} (or \textit{ultra luminous} state) have been shown by only a few sources (e.g. GRO J1655-40 and XTE J1550-564, see e.g. \cite{Motta2012} and \cite{Motta2014b}). This state can be compared to the SIMS, even though it is characterized by significantly higher luminosities. As in the case of the SIMS and the HIMS, both the soft, thermal component from the disk and the hard emission are clearly visible in the spectra seen during these states, while the variability is normally found within 5 and 10\% rms.

\end{itemize}

X-ray binaries, however, also show variability on much shorter times-scales, that cannot be easily studied inspecting a light-curve. Therefore, the Fourier analysis is commonly used to evidence vary fast aperiodic and quasi-periodic variability by producing power-density spectra (PDS). 

In the PDS from BH XRBs we observe several different features, ranging from various types of broad-band noise spanning several decades in frequency (i.e. essentially scale-free), to much more narrow features: the so-called quasi periodic oscillations (QPOs).
QPOs have been observed in practically all kinds of accreting systems (in CVs, XRBs, ULXs and in AGNs). 
These peaks yield accurate centroid frequencies that can be associated with motion and/or accretion-related timescales. The study of variability in general, and QPOs in particular, provides a way to explore the accretion flow around BHs in ways which are inaccessible via energy spectra alone. The association of QPOs with specific spectral states and transitions suggests that they could be a key ingredient in understanding the underlying physical conditions that produce these states. Furthermore, being produced in the vicinity of relativistic objects such as BHs and NSs, they are expected to carry information about the condition of matter in the strong field regime. Therefore, understanding them is key to use QPOs as powerful tools to test the predictions of the Theory of General Relativity.

\section{Quasi Periodic Oscillations in BH XRBs}

QPOs in BH and NS XRBs have been known for many years and have been divided in various classes. QPOs in BH XRBs are normally divided into two large groups, based on the frequency range where they are usually detected: the low frequency QPOs and the high-frequency QPOs. The former are observed below $\sim$50 Hz, the latter are normally found above $\sim$100Hz (but see the case of GRS1915+105, \cite{Belloni2012}) and up to $\sim$500Hz.

\begin{figure*}[t]
\includegraphics[width=0.95\textwidth]{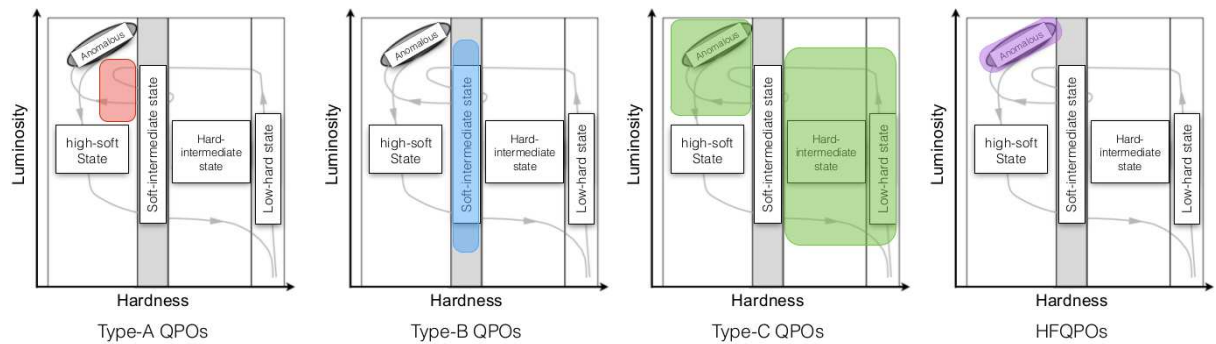}
\caption{Typical black hole transient hardness Intensity diagram pattern. The q-shaped track is divided in five spectral timing states and the location of the different types of low frequency and high-frequency QPOs is marked.}
\label{fig:HID}
\end{figure*}

\begin{figure}[t]
\includegraphics[width=0.45\textwidth]{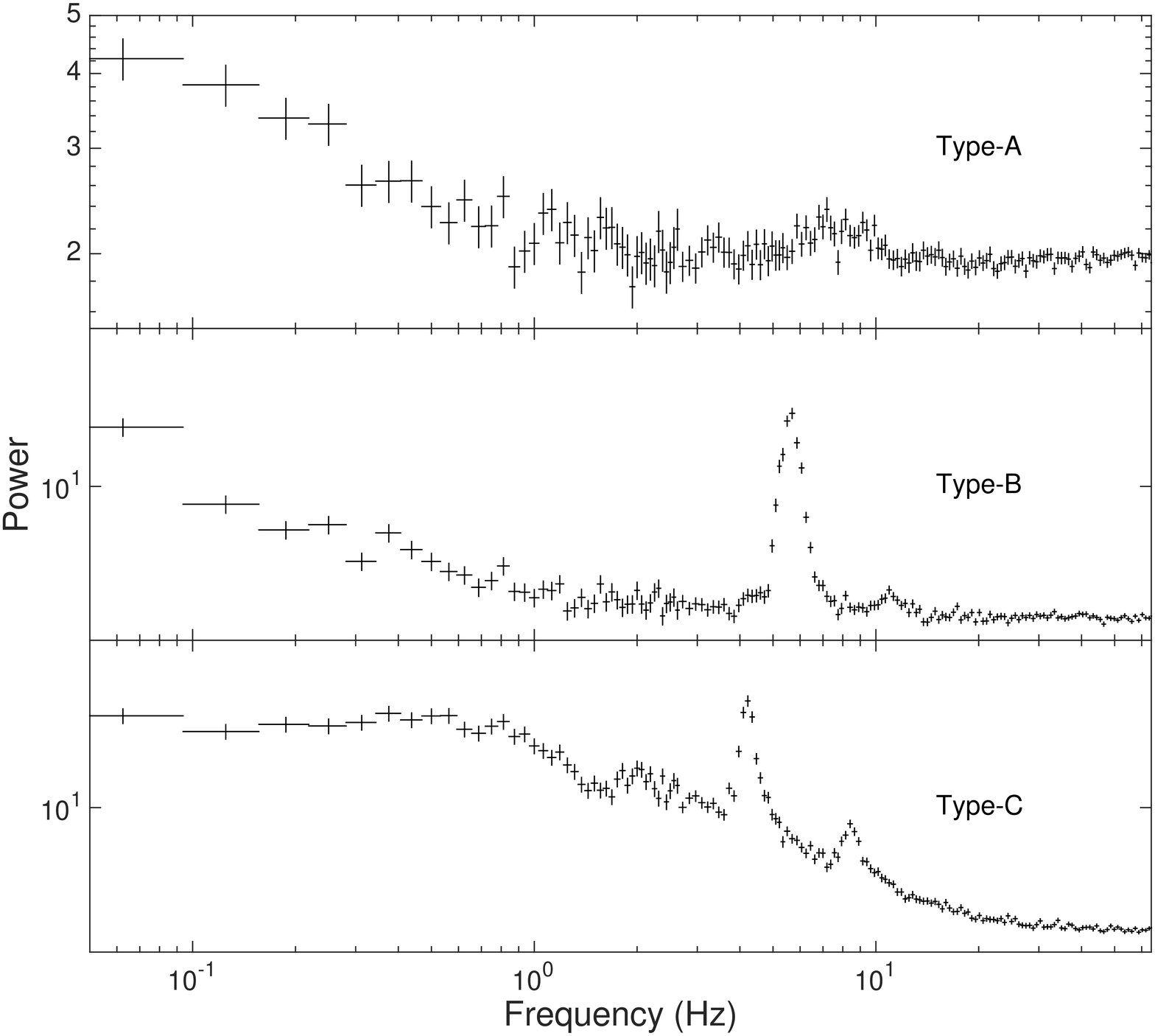}
\caption{Examples of type A, B and C QPOs from our GX 339-4 observations. The centroid peak is indicated with an arrow. The contribution of the Poisson noise was not subtracted. Taken from \cite{Motta2011a}.}
\label{fig:LFQPOs}
\end{figure}

\subsection{Low Frequency QPOs}

Low-frequency QPOs (LFQPOs) with frequencies ranging from a few mHz to $\sim$30 Hz are common features in almost all black transient BHBs and were already found in several sources with \emph{Ginga} and divided into different classes  (see e.g. \cite{Miyamoto1991} for the case of GX 339-4 and \cite{Takizawa1997} for the case of GS 1124-68). Observations performed with the Rossi X-ray Timing Explorer (RXTE) have led to an extraordinary progress in our knowledge on properties of the variability in BHBs (see \cite{VDK2006}, \cite{Remillard2006}, \cite{Belloni2011}) and it was only after RXTE was launched that LFQPOs were detected in most observed BHBs (see \cite{VDK2004}. 

Three main types of LFQPOs, dubbed types A, B, and C, originally identified in the Power Density Spectra (PDS) of XTE J1550-564 (see \cite{Wijnands1999}; \cite{Homan2001}; \cite{Remillard2002}), have been seen in several sources.  

The different types of QPOs are currently identified on the basis of their intrinsic properties (mainly centroid frequency and width, but energy dependence and phase lags as well), of the underlying broad-band noise components (noise shape and total variability level) and of the relations among these quantities.

\paragraph{Type-A QPOs} 

Type-A QPOs (see Fig.  \ref{fig:LFQPOs}, top panel and \ref{fig:HID}, left panel) are the less common type of QPOs in BHBs. In the entire RXTE archive only about 10 type-A QPOs have been found. They normally appear in the HSS, just after the hard to soft transition has taken place, when the overall variability is already quite low. They usually appear close in time to the type-B QPOs. \\
Type-A QPOs (Fig.  \ref{fig:LFQPOs}, top panel) are characterized by a weak (few percent rms) and broad ($\nu/\Delta\nu$ $\leq$3) peak around 6-8 Hz. Neither a subharmonic nor a second harmonic are usually present (possibly because of the width of the fundamental peak), whereas a very low amplitude red noise is associated with type-A QPOs. Originally, these LFQPOs were dubbed \textit{type A-II} by \cite{Homan2001}. LFQPOs dubbed \textit{type A-I} (\cite{Wijnands1999}) were strong, broad and associated with a very low-amplitude red noise. A shoulder on the right-hand side of this QPO was clearly visible and interpreted as a very broadened second harmonic peak. \cite{Casella2005} showed that this \textit{type A-I} LFQPOs should be classified as a Type-B QPOs. \\
Type-A QPOs have been associated to the flaring branch oscillations (FBOs) seen in NS low-mass X-ray binaries (see \cite{Casella2005}), but the low number of detections in both BHs and NSs prevents a secure association between the two classes. For the same reason, an explanation for the existence of Type-A QPOs is difficult. \cite{Tagger1999} proposed a model based on the \textit{accretion ejection instability} (AEI), according to which a spiral density wave in the disc, driven by magnetic stresses, becomes unstable by exchanging angular momentum with a Rossby vortex. This instability forms low azimuthal wavenumbers, standing spiral patterns which would be the origin of LFQPOs. 
\cite{Varni`ere2002} and \cite{Varni`ere2012} suggested that type-A QPOs could be produced through the AEI in a relativistic regime, where the AEI coexist with the Rossby Wave Instability (RWI) (see \cite{Tagger1999}).

\paragraph{Type-B QPOs} 

Type-B QPOs  have been seen in a several BHBs and they appear during the SIMS, which is essentially defined on the presence of this QPO type. \\
Type-B QPOs (Fig.  \ref{fig:LFQPOs}, middle panel and Fig. \ref{fig:HID}) are characterized by a relatively strong ($\sim$4\% rms) and narrow ($\nu/\Delta\nu$ $\geq$6) peak, which is found in a narrow range of centroid frequencies, i.e.  around 6 Hz or 1-3 Hz (\cite{Motta2011}). A weak red noise (few percent rms or less) is detected at very low frequencies ($\leq$0.1 Hz). A weak second harmonic is often present, sometimes together with a subharmonic peak. In a few cases, the subharmonic peak is higher and narrower, resulting in a \textit{cathedral-like} QPO shape (see \cite{Casella2004}). 
Rapid transitions in which type B LFQPOs appear/disappear are often observed in some sources (e.g. \cite{Nespoli2003}). These transitions are difficult to resolve at present, as they take place on a timescale shorter than a few seconds. \\
It is worth noticing that type-B QPOs have been associated to the normal branch oscillations (NBOs) seen in NSs and that both type-B QPOs and NBOs appear at about 6 Hz despite the obvious difference in the mass distributions of the host systems.\\
Type-B QPOs have been associated to the relativistic jets usually seen in X-ray binaries (\cite{Fender2009}) in transition from hard to soft state. \cite{Motta2015} has shown that the intrinsic power of type-B QPOs is higher for low inclination sources (i.e. for sources where the angle between the line of sight and the accretion disk axis is small). This property support the hypothesis that type-B QPOs are related to the relativistic jet (\cite{Fender2009}), since there is no other obvious mechanism that would be stronger face-on. Among the attempts to explain the origin of type-B QPOs, the model proposed by \cite{Varni`ere2002} and \cite{Varni`ere2012} is noteworthy: here the type-B QPOs would arise from the AEI, but differently from type-A and type-C QPOs, they would be produced in a semi-relativistic regime.

\paragraph{Type-C QPOs}

Type-C QPOs are by far the most common type of QPO in BHBs. They can be detected pretty much in any spectral state (see \cite{Motta2012}): they are commonly observed in the LHS and in the HIMS, where their frequency ranges between few mHz and about 10 Hz, but also in the HSS and in the ULS (see e.g. \cite{Motta2012}), where they can reach $\sim$30 Hz. \\
Type-C QPOs (Fig.  \ref{fig:LFQPOs}, bottom panel and Fig. \ref{fig:HID}) are characterized by a
strong (up to 20\% rms), narrow ($\nu/\Delta\nu$ $\geq$10) and variable peak (its centroid frequency and intensity varying by several percent in a few days; see, e.g., \cite{Motta2015}) superposed on a flat-top noise  that steepens above a frequency comparable to the QPO frequency. A subharmonic, a second harmonic peak are often seen and sometimes even a third harmonic peak.  The total (QPO plus flat-top noise) fractional rms variability can be as high as 40\%. The frequency of the type-C QPOs correlates both with the flat-top noise break-frequency (\cite{Wijnands1999a} and with the characteristic frequency of some broad components seen in the PDS at higher frequency ($>$20Hz, see \cite{Psaltis1999}). Type-C QPOs have been associated to the horizontal branch oscillations (HBOs) seen in NS, that also show significant variations in frequency, easily reaching 50-100 Hz.\\
Differently from type-A and type-B QPOs,there are several models attempting to explain the origin of type-C QPOs.  These models are  based on two different mechanisms: instabilities and geometrical effects. In the latter case, the physical process typically invoked is precession. 

\begin{itemize}
\item{\textbf{Instabilities: }} \cite{Titarchuk2004} proposed the so called \textit{transition layer model}, where type-C QPOs are the result of viscous magneto-acoustic oscillations of a spherical bounded transition layer, formed by matter from the accretion disc adjusting to the sub-keplerian boundary conditions near the central compact object.
\cite{Cabanac2010} proposed a model to explain simultaneously type-C QPOs and the associated broad band noise. Magneto-acoustic waves propagating within the corona makes it oscillate, causing a modulation in the efficiency of the Comptonization process on the embedded photons. This should produce both the type-C QPOs (thanks to a resonance effect) and the noise that comes with them.
In the framework of the AEI, type-C QPOs would be produced in the non-relativistic regime, where the RWI does not play any significant role (see  \cite{Varni`ere2002} and \cite{Varni`ere2012}). 

\item{\textbf{Geometrical effects:} } \cite{Ingram2009} proposed a model based on the relativistic precession as predicted by the theory of General Relativity that attempts to explain type-C QPOs and their associated noise. 
This model requires a cool optically thick, geometrically thin accretion disc  (\cite{Shakura1973}) truncated at some radius, filled by a hot, geometrically thick accretion flow. This geometry is known as \textit{truncated disc model} (\cite{Esin1997}, \cite{Poutanen1997}). 
In this framework, type-C QPOs arise from the Lense-Thirring precession of a radially extended section of the hot inner flow that modulates the X-ray flux through a combination of self-occultation, projected area and relativistic effects that become stronger with inclination (see \cite{Ingram2009}).  
The broad-band noise associated with type-C QPOs, instead, would arise from variations in mass accretion rate from the outer regions of the accretion flow that propagate towards the central compact object, modulating the variations from the inner regions and, consequently, modulating also the radiation in an inclination-independent manner (see \cite{Ingram2013}). 

\end{itemize}

\subsection{High Frequency QPOs}

Among the most important discoveries that RXTE allowed there is the detection of  the so-called kHz QPOs
in NS binaries (see \cite{VDK2006}). This result opened a window onto high-frequency
phenomena in BHBs. The first observations of the very bright system GRS
1915+105 led to the discovery of a transient oscillation at $\sim$67 Hz (\cite{Morgan1997}), the first high-frequency QPO (HFQPO) in a BHB. Since then, sixteen years of RXTE observations have yielded only
a handful of detections in other sources (XTE J1550-564, GRO J1655-40, XTE J1859+226, H 1743-322 , GX 339-4, XTE J1752-223, 4U 1630-47, GRS J1915+105, IGR J17091-3624), although GRS 1915+105 seems to be an
exception, with a remarkably high number of detected high-frequency QPOs (see e.g. \cite{Belloni2012}).

\begin{figure}[t]
\includegraphics[width=0.45\textwidth]{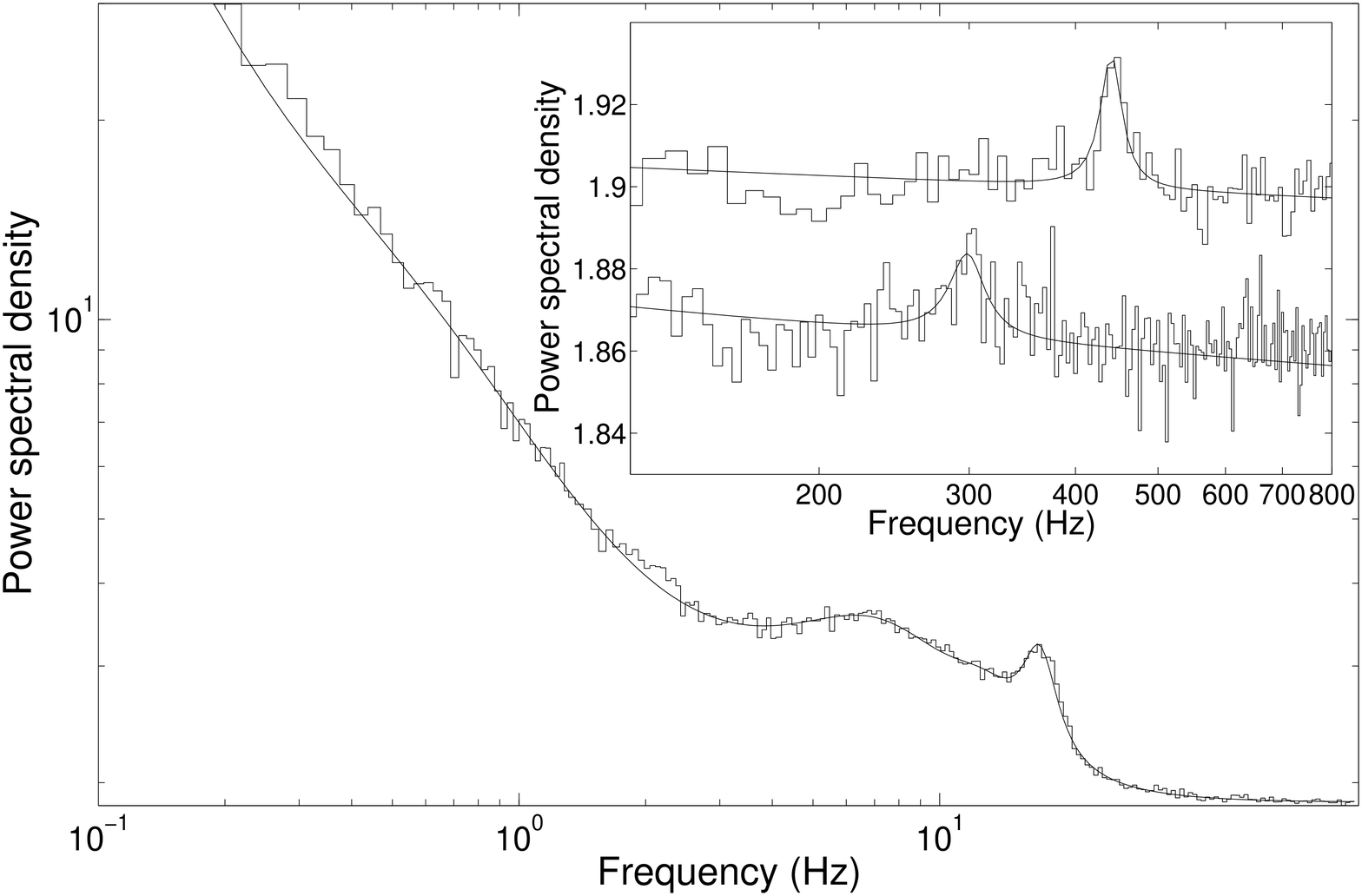}
\caption{Power spectrum of GRO J1655-40 displaying three simultaneous QPO peaks (marked
by arrows): the type C at $\sim$18 Hz, upper and lower high frequency QPO at $\sim$300 and
$\sim$450 Hz, respectively. From \cite{Motta2014}.}
\label{fig:HFQPOs}
\end{figure}

The properties of the few confirmed HFQPOs (\cite{Belloni2012}) can be summarized as follow:

\begin{itemize}

\item They appear only in observations at high flux/accretion rate (see Fig. \ref{fig:HID}). This might 
at least partly due to a selection effect, but not all high-flux observations lead to
the detection of a HFQPO, all else being equal, indicating that the properties of
these oscillations can vary substantially even when all other observables do not
change.

\item They can be observed as single or double peaks. Only one source, GRS J1655-40 (see Fig. \ref{fig:HFQPOs}), showed two clear simultaneous peaks
(\cite{Strohmayer2001}, \cite{Motta2014}), while all the others only showed single peaks, sometimes at different frequencies (see Tab. 1 in \cite{Belloni2014}. 
In XTE J1550-564, the two detected peaks (\cite{Remillard2002}) have been detected simultaneously after averaging a number of observations, but the lower one with a 2.3 $\sigma$ significance
 (\cite{Miller2001}). \cite{Mendez2013},
on the basis of their phase lags, suggested that the two detected peaks might be the
same physical signal at two different frequencies. H 1743-322 showed a clear HFQPO and a weak second simultaneous peak (\cite{Homan2005}).
A systematic analysis of the data from GRS 1915+105 (\cite{Belloni2013}) led to the detection of 51 HFQPOs, most of which at a centroid frequency between 63
and 71 Hz. All detections corresponded to a very limited
range in spectral parameters, as measured through hardness ratios. Additional peaks at 27, 34 and 41 Hz were discovered by \cite{Strohmayer2001a}, \cite{Belloni2001} and \cite{Belloni2013}. The most recent HFQPO discovered, in IGR
J170913624, is consistent with the average frequency of the 67 Hz QPO in GRS
1915+105 (\cite{Altamirano2012}).

\item Typical fractional rms for HFQPOs are 0.5-6\% increasing steeply with energy, in
the case of GRS 1915+105 reaching more than 19\% at 20-40 keV (see right panel
in Fig. 6 of \cite{Morgan1997}). Quality factors Q\footnote{Q is defined as the
ratio between centroid frequency and FWHM of the QPO peak.} are around 5 for
the lower peak and 10 for the upper. In GRS 1915+105, a typical Q of $\sim$20 is
observed, but values as low as 5 and as high as 30 are observed, too.

\item Time lags of HFQPOs have been studied for four sources (\cite{Mendez2013}). The
lag spectra of the 67 Hz QPO in GRS 1915+105 and IGR J170913624 and of the
450 Hz QPO in GRO J1655-40 are hard (hard photons variations lag soft photons
variations), while those of the 35 Hz QPO in GRS 1915+105 are soft. The 300 Hz
QPO in GRO J1655-40 and both HFQPOs in XTE J1550-564 are consistent with
zero (suggesting that the two HFQPOs in XTE J1550 are the same feature seen at different frequencies). 

\item For three sources, GRO J1655-40, XTE J1550-564 and XTE J1743-322, the two
observed frequencies are close to being in a 3:2 ratio (\cite{Strohmayer2001}, \cite{Remillard2002}, \cite{Remillard2006}), which has led to a family of models, known as  \textit{resonance models} (see e.g. \cite{Abramowicz2001}). For GRS 1915+105 the 67 Hz and 41 Hz QPOs,
observed simultaneously, are roughly in 5:3 ratio. The 27 Hz would correspond to
2 in this sequence.

\end{itemize}

\subsubsection{Models for HFQPOs}\label{sec:hfqpos_models}

So far many models have been proposed to describe HFQPOs of BHBs, all involving in some form the predictions of the Theory of General relativity. 

\begin{itemize}

\item The relativistic precession model (RPM), was originally proposed by \cite{Stella1998}, \cite{Stella1999}  to explain the origin and the behaviour of the LFQPO and kHz QPOs in NS X-ray binaries and later extended to BHs (\cite{Stella1999a},\cite{Motta2014},\cite{Motta2014a}). The RPM associates three types of QPOs observable in the PDS of accreting compact objects  to a combination of the fundamental frequencies of particle motion. The nodal precession frequency (or Lense-Thirring frequency) is associated with Type-C QPOs LFQPOs, while the periastron precession frequency and the orbital frequency are associated with the lower and upper HFQPO, respectively (or to the lower and upper kHz QPO in the case of NSs). In this incarnation, through three relatively simple equations, the RPM connects the frequencies of a certain set of QPOs to the mass and spin of the central BH and the radius where the QPOs are produced. Therefore, under precise conditions, the RPM can be used to inferm the mass and the spin of the BH around which QPOs arise (see \cite{Motta2014a} and  \cite{Motta2014b}).
The relativistic precession model has been proposed in two other versions. In \cite{Bursa2005} it is assumed that radiation is modulated by the vertical oscillations of a slightly eccentric fluid slender torus formed close to the ISCO. Stuchlik and collaborators proposed a further version of the relativistic precession model, that has been studied in many papers by this group. Here the model is related to the warped-disk oscillations discussed by \cite{Kato2004} (see below).

\item The the warped disc model proposed by \cite{Kato2004}, \cite{Kato2004a}, states that the HFQPOs are resonantly excited by specific disc deformations – warps. The model was generalized to include precession of the warped disk in \cite{Kato2005a} and spin-induced perturbations were included in  \cite{Kato2005a}.

\item \cite{Abramowicz2001} and \cite{Kluzniak2001} introduced the nonlinear \textit{resonance model}, which was later   studied extensively by them as well as by other authors. This model is based on the assumption that non-linear 1:2, 1:3 or 2:3 resonance between orbital and radial epicyclic motion could produce the HFQPOs observed in both BH and NS binaries. Later on, \cite{Abramowicz2004} proposed another version of this model, called  the Keplerian non-linear resonance model, where the resonance occurs between the radial epicyclic frequency and the orbital frequency instead of between the radial epicyclic frequency and the vertical frequency. 
These \textit{resonance models}  successfully explain
black hole QPOs with frequency ratio consistent with 2:3 or 1:2 (see Sect. 3.2). As a
given resonance condition is verified only at a fixed radius in the disk, the QPO frequencies
are expected to remain constant, or jump from one resonance to another.

\end{itemize}

\section{Conclusions}

After almost 25 years from their discovery, the origin of QPOs in BHBs is still unclear and there is no consensus about their physical nature. However, it is now clear the study of variability in general, and QPOs in particular, provides a way to explore the accretion flow around BHs in ways which are inaccessible via energy spectra alone.
We are now starting to comprehend the fundamentally important physical information hidden in the fast variability properties of accreting systems. The enormous amount of data acquired so far and and soon to be available thanks to the currently flying and future X-ray missions (e.g. XMM-Newton, \textit{Swift}, Astrosat and Astro-H) provide the most desirable workbench to test the growing number of theoretical models attempting to explain the fast variability typical of accreting compact objects.

The knowledge that we obtained over the last decades now allows us to fully exploit the potentialities of timing as a powerful diagnostic tools, especially when used in association with the spectral analysis, in order to unveil the secrets of accretion and of the accretion-ejection mechanism, and, ultimately, of the effects of gravity in extremes conditions. 

\acknowledgements
SEM acknowledges acknowledges support from the ESAC Faculty and the Glasstone Fellowship program.

\newpage%%%%%%%%%%%%%%%%%%%%%%%%%%%%%%%%%%%%%%%%%%%%%%%%%%%%%%

\bibliographystyle{apalike}
\bibliography{biblio}

\end{document}